\documentstyle[12pt,prepstyle]{article}
\begin{document}
\title{Canonical quantum gravity}
\author{Karel V. Kucha\v{r}}
\affil{Department of Physics, University of Utah,\\
 Salt Lake City,
             Utah 84112, U.S.A.}

\beginabstract
This is a review of the aspirations and disappointments of the
canonical quantization of geometry. I compare the two chief ways of
looking at canonical gravity, geometrodynamics and connection
dynamics. I capture as much of the classical theory as I can by
pictorial visualization. Algebraic aspects dominate my description of
the quantization program. I address the problem of observables. The
reader is encouraged to follow the broad outlines and not worry about
the technical details.
\endabstract

\section*{CLASSICAL CANONICAL GRAVITY}

\section*{Dynamical laws and instantaneous laws} One of the main
preoccupations of classical physics has been finding the laws
governing physical data. One of the oldest schemes of quantization has
been to subject such data to canonical commutation relations.  I am
going to review where this program leads us when it is applied to
geometry.

     Classical physics deals with two kinds of laws: dynamical laws,
and instantaneous laws. The discovery of dynamical laws started the
Newtonian revolution. The first instantaneous law was found by Gauss:
at any instant of time, the divergence of the electric field is
determined by the distribution of charges. In empty space, the
instantaneous electric field is divergence-free.

\section*{Theorema egregium} Without knowing it, and without most of
us viewing it this way, Gauss also came across the fundamental
instantaneous law of general relativity: the Hamiltonian constraint.
This constraint is a simple reinterpretation of the famous result
Gauss obtained when studying curved surfaces embedded in a flat
Euclidean space \cite{Gauss}.

     To start with, Gauss drew the key distinction between intrinsic
and extrinsic properties of a surface. The intrinsic properties are
not changed by bending the surface without stretching; the extrinsic
ones are. The basic intrinsic property of a surface is its {\em
intrinsic metric\/}; the basic extrinsic property, its {\em extrinsic
curvature\/}. These are highlighted on the umbrellas of Figure~1.
  \begin{figure}
  \vspace*{4.0in}
  \caption{Intrinsic metric and extrinsic curvature.}
  \end{figure}
The network of distances from the tip of the umbrella along the ribs
is encapsulated by the familiar metric tensor
\begin{equation}
   \d s^{2} = g_{ab}(x)\,\d x^{a}\d x^{b} .
\end{equation}
The ribs lie in the planes passing through the shaft of the umbrella;
they represent its {\em normal sections\/}. Each normal section has a
radius of curvature, $r$, whose reciprocal value, $k$, is the
curvature of the normal section. The extrinsic curvature, $K_{ab}$, of
the surface is an inventory of the curvatures of all its normal
sections:
\begin{equation}
           r^{-1} = k = K_{ab}(x)\, \d x^{a} \d x^{b}\ / \ \d s^{2} .
\end{equation}

{}From the metric, one can derive other intrinsic objects: lengths,
angles, and areas. Geodesics are completely determined by the
metric. So is the {\em parallel transport} of a tangent vector: a
vector is parallel transported from a point to a neighboring point
if it keeps its angle with the geodesic segment connecting the points.
In its turn, parallel transport leads to the concept of {\em scalar}
(or Gaussian) {\em curvature} (Figure 2):

\begin{figure}
                          \vspace*{4.0in}
                          \caption{Scalar curvature.}
\end{figure}

     Take a curve enclosing the tip of the umbrella. Mark where you
want to {\bf START}, take the tangent vector to the curve, and
parallel transport it along the curve back to the starting point. On
the way, the tangent vector (bold) rotates clockwise with respect to
the parallel transported vector (double arrow). If the umbrella were a
plane (I would not like to use such an umbrella on a rainy day), the
tangent vector would run all around the clock. On the bulging umbrella
of Figure~2, it does not quite make it; it still has an angle
$\delta\omega$ to go. As the curve is drawn tighter and tighter around
the tip, the deficit angle $\delta\omega$ becomes proportional to the
surface area $\delta\Sigma$ surrounded by the curve.  The ratio of
these two quantities defines the scalar curvature:
                               \begin{equation}
                       \delta\omega = \case{1}{2}R \,\delta\Sigma \, .
                               \end{equation}
(Gauss did not obtain the scalar curvature this way; by brute force,
he expressed it as a function of the metric and its first and
second derivatives: $R[g]$.)

     Let the ribs again represent normal sections. Of all the ribs,
one has the gentlest bending, $k_{\rm MIN}$, and another has the
steepest bending, $k_{\rm MAX}$. Which rib is which, how large is $k_{\rm
MAX}$, and how small is $k_{\rm MIN}$, depends on the wind. On a quiet
day, all ribs have the same curvature $k_{\rm MAX} = k_{\rm MIN}$. The
values $k_{\rm MAX}$ and $ k_{\rm MIN}$ are called {\em principal
curvatures\/}.

     Because the principal curvatures depend on the wind, they are
clearly extrinsic rather than intrinsic properties of an umbrella.
However, their {\em product\/}, called the {\em total curvature\/},
remains always the same. Indeed, it is proportional to the scalar
curvature which, as we have seen, is an intrinsic property of the
umbrella:
                                \begin{eqnarray}
              R & = & 2 \ \ k_{\rm MAX} \cdot k_{\rm MIN} \nonumber \\~
      {\bf scalar \  curvature} & = & 2 \ \ {\bf total\  curvature}\,.
                                \end{eqnarray}
Gauss has found many remarkable theorems in his life, but this one he
himself regarded to be remarkable: he named the result (4) {\em
theorema egregium\/}.

     By looking at the surface of a single umbrella, we cannot be sure
if it is worn by an upright old man walking across Great Plains, or if
it protects a crooked old goblin wedged into a curved space.  However,
if the intrinsic metric and extrinsic curvature of all possible
umbrellas are connected by Gauss' theorema egregium, we can safely
conclude that the space in which they are embedded is flat and
Euclidean. The flatness of space is thereby guaranteed by the match of
certain intrinsic and extrinsic properties of all embedded surfaces.

     Still, what has all of this to do with the assertion that
dynamics is a consequence of instantaneous laws? There are no instants
and hence no dynamics in a Euclidean space. To talk about instants,
space must be Lorentzian. In a flat three-dimensional Lorentzian
spacetime the theorema egregium still holds. The only thing we need to
change is the sign. On every spacelike surface,
       \begin{equation}
R = -2\, k_{\rm MAX} \cdot k_{\rm MIN}\,.\ \ \  {\rm (Lorentzian)}
       \end{equation}
And, conversely, if the Lorentzian theorema egregium (5) holds on
every spacelike surface in a three-dimensional Lorentzian spacetime,
we can be sure that the spacetime is flat.

     Now, the Lorentzian theorema egregium is an instantaneous law.
On the other hand, the statement that spacetime is flat is a dynamical
law, albeit a very simple dynamical law, about geometry. Roughly
speaking, it tells us that there is no dynamics: spacetime remains flat all
the time. This argument illustrates how an instantaneous law, the
Lorentzian theorema egregium, can lead to a dynamical law, that the
spacetime is flat.

     General relativity, I remember someone saying, does not confine
us to Euclidean barracks. Even if the spacetime is empty (which, for
simplicity, I shall assume for the rest of my lecture), its dynamics
is quite rich. The ripples of gravitational radiation can travel
around, interfere, attract each other, and amplify. They can hold
themselves together in a gravitational geon. Part of the gravitational
radiation can leak out, part of it may collapse and form a black hole.
I find it quite surprising that all this dynamics is encoded in an
almost trivial generalization of Gauss' theorema egregium:

     The intrinsic geometry and the extrinsic curvature of a
three-dimensional hypersurface embedded in a four-dimensional
Riemannian spacetime have the same definition and the same geometric
significance as those of a two-dimensional surface in a
three-dimensional flat space. However, instead of two principal
sections there are three, with principal (extremal) curvatures
$k_{1}$, $k_{2}$, and $k_{3}$. One cannot define the total curvature $T$
as the product of a selected couple of principal curvatures. As a true
egalitarian, one takes
       \begin{equation}
T = k_{1}k_{2} + k_{1}k_{3} + k_{2} k_{3}\, .
       \end{equation}
Similarly, there is no single surface on which one can determine the
deficit angle $\delta\omega$ by parallel transporting the tangent
vector along a curve. Instead, one chooses three perpendicular
surfaces passing through the tip of a three-dimensional umbrella, and
determines the three deficit angles $\delta\omega_{1}$,
$\delta\omega_{2}$, and $\delta\omega_{3}\,$. The scalar curvature $R$
has the geometric meaning
       \begin{equation}
\case{1}{2} \, R = \frac{\delta\omega_{1}}{\delta\Sigma_{1}} +
                   \frac{\delta\omega_{2}}{\delta\Sigma_{2}} +
                   \frac{\delta\omega_{3}}{\delta\Sigma_{3}}\  .
       \end{equation}
Compare now the total curvature (6) and the scalar curvature (7) of a
hypersurface in an arbitrary Ricci-flat spacetime. Behold, the
theorema egregium still holds:
       \begin{equation}
R = \left\{ \begin{array}{l}
            - \ {\rm (Lorentzian)} \\
            + \ {\rm (Euclidean)}
            \end{array}
            \right.
            \ \ 2T \, .
       \end{equation}
Inversely, if the scalar curvature is related to the total curvature
by Eq.(8) on any spacelike hypersurface, the spacetime is necessarily
Ricci-flat. Therefore, the statement that the theorema egregium holds
at any instant is entirely equivalent to the Einstein law of
gravitation in empty space!

     I am sorry that the continuation of my narrative requires some
juggling of indices. The total curvature (6) is a quadratic
combination of the three principal curvatures. Because each of these
is a linear function of the extrinsic curvature (2), the total
curvature can be expressed as a quadratic form of the extrinsic
curvature:
       \begin{equation}
T = - \case{1}{2} K_{ab} \,G^{ab\,cd} \,K_{cd}\, .
       \end{equation}
The coefficient
 \begin{equation}
G^{ab\,cd} =  \case{1}{2} ( g^{ac}g^{bd} + g^{ad}g^{bc} - 2\,g^{ab}g^{cd} )
       \end{equation}
is called the {\em supermetric\/}. Symmetric pairs of covariant indices
can be raised by the supermetric, and symmetric pairs of contravariant
indices lowered by its inverse, $G_{ab\,cd}\,$. The contravariant
version of the extrinsic curvature is
                                       \begin{equation}
p^{ab}\  \mbox{:=}\  G^{ab\,cd}\,K_{cd} \,  .
                                       \end{equation}
The total curvature is a quadratic form of $p^{ab}$, and the
Lorentzian theorema egregium (8) assumes the form \footnote{The
notation $R(x; \, g]$ emphasizes that R is a function of $x$ and a
functional of $g\,$.}
                               \begin{equation}
H(x)\  \mbox{:=}\  p(x) \cdot G(x;\,g) \cdot p(x) - R(x;\,g] = 0\, .
                               \end{equation}

     The theorema egregium is the most fundamental instantaneous law
of Einstein's theory of gravitation. Gauss did not realize that the
theory of curved surfaces in a flat Euclidean space (and of curved
hypersurfaces in a Ricci-flat spacetime) is subject to yet another
instantaneous law, closely resembling the law which he had found for
electricity. As shown by Codazzi \cite{Codazzi}, the covariant
divergence of the extrinsic curvature $p^{ab}$ vanishes:~\footnote{I
write $\simeq$ whenever I want to sweep a numerical factor under the
rug.}

                               \begin{equation}
H_{a}\  \mbox{:$\simeq$}\  {}_{g}\! \nabla _{b} \, p_{a}{}^{b}(x) = 0 \,.
                               \end{equation}

\section*{Canonical geometrodynamics} The stage is now ready for
stating (not proving, nor even properly explaining) the sea change
which the theorema egregium (12) and the Codazzi law (13) suffered a
century later. Working from quite an opposite direction of variational
principles and Hamiltonian dynamics, Dirac \cite{Dirac} and Arnowitt,
Deser, and Misner \cite{ADM} have shown that the intrinsic metric
$g_{ab}$ and the (densitised) extrinsic curvature $p^{ab}$ are
canonically conjugate to each other.  In canonical theory, the
instantaneous laws (12) and (13) are called the Hamiltonian and
diffeomorphism constraints. They start playing a double role. On one
hand, they {\em restrict\/} the canonical data. On the other hand, as
dynamical variables on the phase space, they become capable of {\em
evolving\/} the canonical data. The Poisson bracket of the data with
$H_{a}(x)$ generates their change by a Lie derivative in the direction
along the hypersurface. Similarly, the Poisson bracket with $H(x)$
generates the change of the data under a normal displacement of the
hypersurface.  These two processes enable us to organize the
embeddings by displacements which deform one embedding into another,
and to correlate the data which the embeddings carry.  Instead of
checking the Einstein law by criss-crossing the spacetime by all
possible hypersurfaces, we obtain it by an orderly Hamiltonian
evolution which smoothly deforms the original hypersurface. The change
of the canonical data by the generators $H_{a}(x)$ and $H(x)$,
together with the statement that the generators, once they generated
the change, are constrained to vanish, {\em is} the Einstein law. This
is the new strange role of the instantaneous laws: they become the
agents of dynamics.

     The change generated by $H_{a}(x)$ is induced by a spatial
diffeomorphism Diff$\Sigma$ on a given hypersurface. This property
gave the Codazzi constraint its new name --- the diffeomorphism
constraint. The constraint ensures that the theory is invariant under
Diff$\Sigma$. In other words, canonical geometrodynamics does not
depend on the intrinsic metric and the extrinsic curvature, but only
on such combinations of these variables which are unaffected by
spatial diffeomorphisms, i.e., only on the intrinsic and extrinsic
geometries.  There are {\em fewer} physical variables than the symbols
which meet the eye.

     This message can also be read backwards: by making the theory
dependent on {\em more} variables, one can make it invariant with
respect to a wider class of transformations. A good example of this
process is triad dynamics.

\section*{Triad dynamics}
 Let us choose as our basic variables a triad $E^{a}_{i},\ i = 1,2,3$,
of orthonormal vectors \cite{triad}. These determine the intrinsic
metric,
            \begin{equation}
g^{ab} = \delta ^{ij} E^{a}_{i} E^{b}_{j}\,,
            \end{equation}
but the metric determines the triad only up to an $x$-dependent SO(3)
rotation. The rotation group SO(3) becomes a gauge group of the
Einstein theory. Canonical analysis reveals that the projected
extrinsic curvature
                     \begin{equation}
-K^{i}_{a}(x) = - K_{ab}(x)E_{j}^{b}\,\delta^{ji}
                        \end{equation}
is the canonical coordinate whose conjugate momentum is the
(densitised) triad $E_{i}^{a}$. The SO(3) rotations of the canonical
variables are generated by the dynamical variable
                                                   \begin{equation}
G_{i}(x)\  \mbox{$:=$}\  \epsilon _{ij}{}^{k}
(-K^{j}_{a}(x))E^{a}_{k}(x) = 0\,,
                                                   \end{equation}
which has the familiar structure of angular momentum. After
generating the rotations, $G_{i}(x)$ is constrained to vanish. The
{\em rotation constraint} (16) ensures that the extrinsic curvature
$K_{ab}$ related to $K_{a}^{i}$ by Eq.(15) is symmetric.

Any vector, $u^{a}$, can be characterized by its internal components,
{}$u^{i}$, in the orthonormal basis $E^{a}_{i}\,$:
                                                   \begin{equation}
u^{a} = u^{i}E^{a}_{i} .
                                                   \end{equation}
Let us parallel transport the vector $u^{a}$ from $x$ to $x + \d x$;
we get the double-arrow vector of Figure~3.
             \begin{figure}
\vspace*{3.5in}
\caption{The SO(3) parallel transport.}
             \end{figure}
There is a basis, $E^{a}_{i}(x + \d x)$, sitting at $x + \d x$ . In
this basis, I draw a vector which has the same components, $u^{i}$, as
the original vector had at $x$. I call it the {\em reproduced
vector\/}.  To turn the reproduced vector into the parallel
transported vector, I must rotate its internal components by an angle
$\delta\omega^{i}$. This angle is a linear function of the
displacement $\d x^{a}$:
                                                   \begin{equation}
\delta\omega^{i} = - \Gamma^{i}_{a}\d x^{a}.
                                                   \end{equation}
The coefficient $\Gamma_{a}^{i}$ tells us how the parallel transport
of a vector affects its internal components. It can be expressed in
terms of the triad $E_{i}^{a}(x)$ and its first derivatives. It is
called the SO(3) connection.

     As usual, the curvature tensor of a connection is defined by the
parallel transport of a vector $u$ along a small parallelogram with the
edges $\d x$ and $\delta x$ (Figure~4).
                               \begin{figure}
\vspace*{3.5in}
\caption{The SO(3) curvature tensor and the box identity.}
                               \end{figure}
The parallel transported vector (shown as double arrow) does not
return back to its original position (shown in bold). To turn the
original vector into the parallel transported vector, we must subject
its internal components to a rotation:
                                    \begin{equation}
\delta\omega^{i} = - R_{ab}{}^{i} \d x^{a} \delta x^{b}.
                                    \end{equation}
The coefficient $R_{ab}{}^{i}$ is the curvature tensor of the SO(3)
connection. It can be expressed in terms of the basis vectors
$E^{a}_{i}\,$, and their first and second derivatives.

     The curvature tensor satisfies the {\em cyclic identity\/}. Its
geometric significance is illustrated on the right in Figure~4.  Take
a small box with edges $u,\,v$ and $w$. Parallel transport $w$ along
the boundary of the face $u,\,v$. The parallel transported vector does
not coincide with $w$; it differs from it by $\delta w$. Repeat this
procedure for the remaining two vectors, and obtain the differences
$\delta u$ and $\delta w$. While none of them in general vanishes,
their sum is identically equal to zero: $\delta u + \delta v + \delta
w \equiv 0\,$.

     It is easy to write down what this geometric construction yields
for a small box whose edges lie in the direction of the orthonormal
vectors $E_{i}^{a}\,$. We obtain
                                                    \begin{equation}
R_{ab}{}^{i}E_{i}^{b} \equiv 0 \,.
                                                    \end{equation}
The SO(3) curvature tensor $R_{ab}{}^{i}[E]$ necessarily satisfies
the box identity (20).

     Once we know the curvature tensor, we can determine the curvature
scalar as in Eq.(7). We take three mutually perpendicular curves, each
of them enclosing a unit area, parallel transport their tangent vectors,
determine the deficit angles, and add them together. In particular, we
can choose for the curves the parallelograms spanned by the pairs of
the orthonormal vectors $E^{a}_{i}$. In this way we learn that
                                                    \begin{equation}
R[E] = -  R_{ab}{}^{i}\, \epsilon _{i}{}^{jk}E_{j}^{a} E_{k}^{b}\,.
                                                    \end{equation}
Algebraically, $R[E]$ can be obtained by substituting the metric (14)
into $R[g]$.

     We can now take the Hamiltonian constraint (12) and express it in
terms of the new canonical variables $-K^{i}_{a}$ and $E_{i}^{a}$. By
using Eqs.(9)--(10) and (14), we cast the total curvature into a form
in which it is quadratic both in $-K^{i}_{a}$ and in $E_{i}^{a}$. The
curvature scalar is a concomitant (21) of the triad $E_{i}^{a}\,$. As
a result, $H$ is set forth as a functional of $-K^{i}_{a}$ and
$E^{a}_{i}$. The diffeomorphism constraint can be handled in the same
way. Neither of the constraints looks any simpler in the new variables
than it did in the old ones.

In addition to the old constraints, we have the rotation constraint
(16). More variables call for more constraints. There are nine entries
in the triad $E^{a}_{i}\,$, while there are only six entries in the
symmetric metric $g_{ab}\,$. Similarly, there are nine entries in
$-K^{i}_{a}\,$, while there are only six in the extrinsic curvature
$K_{ab}\,$. However, the physics depends only on the old set of
variables, $g_{ab}$ and $K_{ab}\,$. The triad $E^{a}_{i}$ enters into
the Hamiltonian and diffeomorphism constraints only through the
combination (14), and the extrinsic curvature given by Eq.(15) is
forced to be symmetric by the rotation constraint (16). The surplus
dynamics -- rotations of the triad generated by the constraint (16) --
is expendable. In the end, only such quantities which are unaffected
by rotations (like $g_{ab}$ and $K_{ab}$) physically matter. We can
return to them, forget about the rotation constraint, and retrieve
geometrodynamics from triad dynamics. To gain more invariance by
introducing more variables is not a big deal.

\section*{Connection dynamics}

     To simplify the constraints, one must go one step beyond
introducing the triads: one must modify the parallel transport.
Figure~5
                               \begin{figure}
\vspace*{4in}
\caption{New parallel transport.}
                               \end{figure}
shows a three-dimensional umbrella protecting us from a storm of
gravitational waves in a four-dimensional Euclidean Ricci-flat
spacetime. Take a tangent vector to the umbrella (shown as a double
arrow) and parallel transport it from the tip along one of the
principal sections. The transported vector is again shown as a double
arrow.  Viewed from the center of curvature, the arc $\d x$ along
which the vector is transported subtends the angle $k \d x$.  Give now
the parallel transported vector an additional twist by the angle $k \d
x$ about the principal direction. The position of the vector after the
twist defines the new parallel transport.

      The twist does not look quite right in the two-dimensional
sketch of the three-dimensional umbrella. It seems that to rotate the
vector about the rib we must rotate the whole tangent plane, and
destroy thereby its tangential character. To see what is happening, we
must look at the tangent plane through one of the most powerful
instruments ever invented by a theoretical physicist: John A.
Wheeler's dimensional magnifying glass \cite{Wheeler}. Under the
glass, the plane thickens into what it actually is, a
three-dimensional tangent space, and the twist moves the vector along
a cone in this space into its new position. The normal to the umbrella
stays fixed because the twist takes place in the plane perpendicular
to the principal section.

     To transport a vector in an arbitrary direction, we must
decompose the displacement $\d x$ into the three principal directions
and perform the appropriate twists one after another.  The new parallel
transport amounts to a single rotation (18) of the reproduced vector.
The angle of rotation is given by a new SO(3) connection \cite{Sen,Ash}

                               \begin{equation}
A_{a}^{i} = \Gamma^{i}_{a} - K_{a}^{i} \,.\ \ \ \ \ {\rm (Euclidean)}
                               \end{equation}

     {}From  the canonical standpoint, it is remarkable that Eq.(22)
represents a canonical transformation: the new SO(3) connection
$A_{a}^{i}$ is a coordinate canonically conjugate to the momentum
$E^{a}_{i}\,$. (To see that $A^{i}_{a}$ is conjugate to $E^{a}_{i}$
is trivial, because $-K^{i}_{a}$ is conjugate to $E^{a}_{i}$. It is
more difficult to prove that $A^{i}_{a}(x)$ can serve as a field
coordinate, i.e., that it has a vanishing Poisson bracket with
$A^{j}_{b}(x')$.) But why should we ever want to perform the
canonical transformation (22)? Cui prodest?

The new parallel transport leads, by Figure~4 and Eq.(19), to the new
curvature tensor $F_{ab}{}^{i}[A]$. In terms of this tensor, the
instantaneous laws of Euclidean Ricci-flat spacetimes take a
remarkably simple form. The new curvature tensor no longer satisfies
the box identity (20). Instead, the expression $F_{ab}{}^{i}
E^{b}_{i}$ yields the supermomentum \footnote{This time, $ \simeq $
sweeps under the rug not only numerical factors, but also the fact
that the rotation constraint is used in rearranging the diffeomorphism
constraint and the Hamiltonian constraint.}
                               \begin{equation}
H_{a} \simeq F_{ab}{}^{i}[A]E^{b}_{i} \, .
                               \end{equation}
The box equation (20) still holds, but no longer as an identity. It is
now equivalent to the Codazzi law (13) in a Ricci-flat spacetime. Even
more remarkably, the $H$ of Gauss' theorema egregium turns out to be
the scalar curvature (21) of the new parallel transport:
                               \begin{equation}
H \simeq F_{ab}{}^{i}[A] \, \case{1}{2} \epsilon _{i}{}^{jk} E^{a}_{j}
E^{b}_{k} \, .
                               \end{equation}
These striking facts were discovered by Ashtekar \cite{Ash}.  It is
quite tempting to call Eq.(24) Ashtekar's theorema egregium: In a
Ricci-flat spacetime, the scalar curvature of Ashtekar's connection
$A$ vanishes on every hypersurface.

     The new connection $A^{i}_{a}$ defines the new covariant
derivative ${}_{A}D _{a}$ acting on internal indices. In terms of this
derivative, the rotation generator (16) takes the form
                               \begin{equation}
G_{i} = {}_{A} D_{a} \, E^{a}_{i} \, .
                               \end{equation}
(Because $E^{a}_{i}$ is a vector density, $ {}_{A}D_{a}$ does not need
to act on spatial indices to produce a scalar density.)

     These are the good news. Now, for the bad news. The transition
{}from a Euclidean to a Lorentzian spacetime enforces a change of sign in
the Gauss theorema egregium. The Ashtekar theorema egregium can absorb
this change of sign only at the price of introducing a complex SO(3)
connection:
                               \begin{equation}
 A^{i}_{a} = \Gamma ^{i}_{a} - i \, K^{i}_{a} \, .\ \ \ \ \ {\rm (Lorentzian)}
                               \end{equation}
To handle this complication in canonical quantum gravity is not
entirely trivial.

\section*{Dialogue
          concerning the two chief systems of canonical gravity: \\
          geometrodynamics and connection dynamics}

     Geometrodynamics and connection dynamics are the two chief forms
of canonical gravity. Let us pause and compare them before proceeding
with quantization. I suppress the turmoil of indices and highlight the
two structures in a table. Let
                                \begin{quote}
{} $\cdot$ denote contraction in spatial indices,

{} $\circ$ denote contraction in spatial indices {\em and}
internal indices,

{}${}^{\ast}$ denote internal dualization,

                                \end{quote}
and the details take care of themselves. Then

                                \begin{displaymath}
                                    \begin{array}{ccc}
{}                          &   {}    &   {} \\
{}                          &   {}    &    {} \\
{\rm GEOMETRODYNAMICS}      &   {}    &   {\rm CONNECTION\  \ DYNAMICS}  \\
{}                          &   {}    &            {}                  \\
{}                          &   {}    &            {}                  \\

                                  \begin{array}{cc}

{\bf Coordinates}      &      {\bf Momenta} \\
{}                     &            {}       \\
        g              &            p      \\
{}                     &            {}      \\
                                    \begin{array}{c}
{\bf Intrinsic} \\
{\bf metric}    \\
{}                                    \end{array} &
                                    \begin{array}{c}
{\bf Extrinsic} \\
{\bf curvature} \\
{}
                                    \end{array}
                                  \end{array}   &
                                  \begin{array}{c}
{\rm CANONICAL} \\
{\rm VARIABLES} \\
{}                                  \end{array} &
                                  \begin{array}{cc}
{\bf Coordinates}           &           {\bf Momenta}    \\
{}                          &            {}              \\
        A                   &              E             \\
{}                          &            {}               \\
                                    \begin{array}{c}
{\bf SO(3)}       \\
{\bf connection}  \\
{\bf (mixed) }
                                    \end{array} &
                                    \begin{array}{c}
{\bf Triad}       \\
{}                \\
{\bf (intrinsic)}

                                    \end{array}
                                  \end{array} \\
{}                       &  {}  &                   {}      \\
{}                       &  {}  &                   {}      \\
{}              & {\rm GENERATORS\ \  OF}   &       {}      \\
{}              &            {}             &       {}      \\
{}                       &  {}  {}          &       {}      \\
p \cdot G(g) \cdot p\,-\,R[g]&\perp\ {\rm EVOLUTION} &  E \circ {}^{\ast}F[A]
                                                    \circ E  \\
{}                       &   {}             &       {}          \\
{}_{g}\nabla \cdot p     & {\rm Diff}\Sigma &       F[A] \circ E  \\
{}                       &       {}         &       {}              \\
{\bf None}               & {\rm ROTATIONS}  &       {}_{A}D \cdot E  \\
{}                       & {}               &       {}
                                \end{array}
\end{displaymath}

     A comparison of the two columnes brings forward a number of simple
observations:
\begin{enumerate}
\item{{\em Variables and constraints.} Geometrodynamics works with
fewer variables and fewer constraints than connection dynamics.  The
geometrodynamical variables are invariant under triad rotations.}
\item{{\em Connection with gauge theories.} The rotation constraint
makes connection dynamics resemble an SO(3) Yang-Mills theory. In
geometrodynamics, the rotation constraint is eliminated and one works
with SO(3)-invariant canonical variables.}
\item{{\em Dimension.} I discussed the constraints in $3+1$ dimensions.
Geometrodynamics remains virtually the same in any dimension $n + 1,\
n \geq 2 $. The SO(3) connection dynamics is intimately adapted to a
three-dimensional space and it is not easily generalized to $n > 3$.}
\item{{\em Positivity restrictions.} The Cauchy problem works only
if the hypersurfaces are spacelike, i.e., if the induced metric $g$ is
positive definite. In geometrodynamics, this puts a restriction on the
domain of the configuration space. In connection dynamics, the metric
is automatically positive definite, as long as the triad is
non-degenerate. However, even for degenerate triads (leading to
degenerate metrics) the formalism seems to make sense, and it is
viable to lift the non-degeneracy restriction.}
\item{{\em Structure of the Hamiltonian constraint.} In
geometrodynamics, $H$ is a {\em quadratic function} of the momentum $p$.
The supermetric $G(g)$ is ultralocal, and there is a local potential
term $R[g]$. In connection dynamics, the potential is absorbed into
the quadratic term; $H$ is a {\em quadratic form} of the momentum $E$.
However, the supermetric ${}^{\ast}F[A]$ is no longer ultralocal, but
merely local.}
\item{{\em Polynomiality of constraints.} In connection dynamics, all
constraints are low-degree polynomials in the canonical variables $A$
and $E$. It was originally claimed that geometrodynamical constraints
are non-polynomial in the canonical variables $g$ and $p$, but a more
careful look \cite{Tate} reveals that a simple scaling by a
power of det$(g)$ also makes them polynomial. However, they are
polynomials of a rather high degree.}
\item{{\em Reality conditions.} Geometrodynamics works with real
canonical variables on a real phase space.. We have seen that in a
Lorentzian spacetime the Ashtekar variable $A$ is necessarily
complex. This forces one to work with complex canonical variables,
either on a complex or a real phase space. A pair $A$ and $E$ of
canonical variables that satisfy the constraints define a real
Ricci-flat spacetime only if they satisfy the {\em reality conditions}
\begin{equation}
E - \bar{E} = 0\, , \ \ A + \bar{A} = \Gamma[E]\, .
\end{equation}
These conditions are non-polynomial in $E$. People replace
\cite{Ash-re} the reality conditions (27) by somewhat weaker
conditions that are polynomial. A simpler procedure is to scale the
second condition (27) by $[{\rm det}(g)]^{2}$ which makes it
polynomial in $E$ and $A$. Whichever way one proceeds, the polynomial
reality conditions are of a rather high degree.}
\end{enumerate}

{}

     In view of these observations, which scheme is simpler,
geometrodynamics or connection dynamics? Simplicity, of course, is in
the eye of the beholder, and my assessment is quite personal.
\begin{enumerate}
\item{I believe that, on one hand, one should not make too much fuss
about the count of the variables and constraints and, on the other
hand, one should not overemphasize the resemblance between connection
dynamics and the SO(3) gauge theories.}
\item{The SO(3) invariance is a
simple consequence of introducing the redundant variables. The ease
with which such variables are eliminated and connection dynamics
reduced back to geometrodynamics may be an indication that the
achieved SO(3) invariance is not that deep. However, one should not
overlook that the mixing of the extrinsic and intrinsic variables
brings in a true simplification of the constraints prior to the
imposition of the reality conditions.}
\item{ Our space is three-dimensional and a theory which makes an
effective use of this fact is not to be blamed. I view the
simplifications which can be achieved only in three dimensions
speaking for rather than against connection dynamics.}
\item{The positivity restrictions on the metric are quite a nuisance in
quantum theory. The possibility of lifting the non-degeneracy
condition on the triad without endangering the connection dynamics is
a real advantage. However, when listing the achievements of quantum
connection dynamics, one should bear in mind that many of these
correspond to situations in which the triad and hence the metric are
degenerate.}
\item{Trading an ultralocal supermetric and a local potential term for
a local supermetric without any potential is an interesting quid pro
quo. Whether such a trade-off pays off in quantum theory depends quite
heavily on whether it is easier or not to turn the new Hamiltonian
constraint into a well-defined operator.}
\item{A low-degree polynomiality is certainly an asset in
quantizing a classical theory. In this respect, the constraints of
connection dynamics are definitely simpler than the geometrodynamical
ones.}
\item{One should bear in mind, however, that connection dynamics is
sooner or later confronted with the task of implementing the reality
conditions. These conditions are unseemly, being polynomials of such a
high degree as the geometrodynamical constraints. The simplifications
which the connection dynamics achieves may thus be a mere temporary
advantage.}
\end{enumerate}

     The Galilean overtones of my section heading are meant to go
beyond a mere joke.  Eliminating variables or constraints is like
getting rid of epicycles. The presence of a potential term may be
aesthetically repugnant like the use of an equant. Nevertheless, the
fact remains that geometrodynamics and connection dynamics are
entirely equivalent at the classical level, just as the Ptolemaic
and Copernican systems are entirely equivalent at the kinematical
level.  The Copernican system may be aesthetically more pleasing, but
its real power emerges only when one starts asking dynamical
questions.  Similarly, the real power of connection dynamics may
emerge only when one starts quantizing the classical theory. This is
the task we should now discuss.

\section*{CONSTRAINT QUANTIZATION: A PROGRAM}

     Canonical gravity is a system whose dynamics is entirely
generated by constraints. Its quantization and interpretation presents
some special difficulties. The ground rules for quantizing constrained
systems were laid by Dirac \cite{Dirac} and refined over the years.
Every major review of canonical quantum gravity %
\cite{ADM,reviews,Ash-book} %
attempted to list a sequence of steps expected to lead to a
satisfactory theory. People more or less agree about what these steps
are, but they do not know how to implement them: by listing the steps,
they present a mere quantization program. I shall picture seven steps
of a quantization program as seven gateways on a road paved with good
intentions.

\section{Fundamental variables}

     The first step is the selection of {\em fundamental variables}.
These are classical dynamical variables that are to be turned into
operators whose commutator algebra replicates the classical Poisson
algebra. The fundamental variables are expected to span a vector space
$\cal V$ closed under the Poisson brackets ${\{\ ,\ \}}$. The space
$\cal V$ should be {\em complete} in the sense that any dynamical
variable $F$ can be approximated by an element of the free algebra
$\cal A$ over $\cal V$, i.e., expressed as a sum of products of the
elements of $\cal V$.

     In geometrodynamics, $\cal V$ is taken to be a real vector space
spanned by $g$, $p$, and the unit dynamical variable 1. In connection
dynamics, $\cal V$ is taken to be a complex vector space spanned by
$A$, $E$, and 1. The elements $V \in \cal V$ are expected to
be mapped into operators $\hat{V} \in \hat{\cal V}$ in such a way that
                 \begin{equation}
V_{3} = \{ V_{1}, V_{2} \}\ \  \Longrightarrow \ \ \hat{V}_{3} = - i [
\hat{V}_{1}, \hat{V}_{2}] \, .
                 \end{equation}

     In geometrodynamics, the metric $g$ should be positive definite.
The positivity conditions cannot be written as relations in $\cal V$,
and their imposition is quite tricky \cite{Ish+}. Connection dynamics
is fully equivalent to geometrodynamics only for non-degenerate triads
$E$. People working in connection dynamics propose not to impose the
condition that $E$ be non-degenerate.

     Connection dynamics has a difficulty which does not exists in
geometrodynamics: not all elements of the complex vector space $\cal
V$ describe a real spacetime. Ultimately, one must impose the {\em
reality conditions} (27). However, $\cal V$ is not closed under the
complex conjugation. The reality conditions thus cannot be formulated
in $\cal V$ (the old SO(3) connection $\Gamma (E)$ does not lie in
$\cal V$). It was proposed \cite{Ash-book,A+T} that at the level
of representing the fundamental variables by operators one should
simply forget about the reality conditions. These are to be taken care
of much later, during the construction of a Hilbert space.  Conforming
to this view, I return to the issue of reality conditions at the last
of my gates.

\section{Dynamical variables (including constraints)}

     In the next step, one must decide on how to turn an arbitrary
dynamical variable $F[g,p] \in \cal A$ or $F[A,E] \in \cal A$ into an
operator. Such variables typically do not lie in $\cal V$, but they
can be approximated by sums of products of the elements of $\cal V$.
Van Hove \cite{Van Hove} has proved that it is impossible to turn
dynamical variables into operators in such a way that Eq.(28) holds
for all of them. Without the guiding principle (28), the quantization
of dynamical variables is subject to factor-ordering ambiguities. It
is popular to dismiss these as `mere quantum-mechanical corrections'.
I do not share this view. Unless one knows how to factor order
significant dynamical variables, one really does not know how to
construct quantum theory. In a sense, the right factor ordering {\em
is} the quantum theory. If one does not set any rules about factor
ordering, one can turn a classical variable $F(Q,P)$ into any quantum
operator one pleases:

     Let $F(Q,P)$ be a classical dynamical variable, and $G(Q,P)$ any
other dynamical variable (whose dimension is that of $F$ divided by
the action). Fix some factor ordering of $\hat{F} = F(\hat{Q},\hat{P})$
and $\hat{G} = G(\hat{Q}, \hat{P})$, and define the quantum
variable
                 \begin{equation}
\hat{F}' \, \mbox{ :=} \, \hat{F} - i\,[\hat{Q},\hat{P}])\hat{G}
                    =\hat{F} + \hat{G}\, .
                 \end{equation}
The quantum variables $\hat{F}'$ and $\hat{F}$ have the same classical
limit, namely, $F(Q,P)$, and yet they differ by an arbitrary operator
$\hat{G}$. This is what a `mere' factor ordering can do.

     A particular case of dynamical variables in canonical gravity are
the constraints. One should turn them into operators $\hat{H}(x)$, $
\hat{H}_{a}(x)$ (and possibly $\hat{G}_{i}(x)$). In field theory, this
presents a regularization problem. Moreover, in the next step of the
quantization procedure one wants to impose the constraints on the
states. This poses a consistency problem. Both of these problems are
troublesome, but they at least impose severe restrictions on the
factor ordering. On the other hand, very little is known and, even
more remarkably, said about what to do with other dynamical variables.

\section{Representation space $\cal F$}

     The operators representing the dynamical variables are expected
to act on a space of states. One way of choosing this space is to rely
on the Schr\"{o}dinger representation: the canonically conjugare pairs
of fundamental variables are taken as multiplication and
differentiation operators acting on functionals of the configuration
variables. Thus, in geometrodynamics $\cal F$ is taken to be a complex
vector space whose elements are the functionals $\Psi [g]$ of the
metric. In connection dynamics, the elements of $\cal F$ are the
functionals $\Psi [A]$ of the Ashtekar connection. \footnote{The
connection representation $\Psi [A]$ is formally related to the
{\em loop representation}. This is discussed by Smolin in this volume,
and in a recent review by Rovelli \cite{Rov-loops}.}

     There is an important difference between the Schr\"{o}dinger
representation for unconstrained systems and the Schr\"{o}dinger
representation in canonical gravity: the representation space ${\cal F}$
is not necessarily assumed to be a Hilbert space, and (real) dynamical
variables are not required to be represented by self-adjoint
operators \cite{K-Ox}. Physically, the Hilbert space structure is
needed to calculate the expectation values of observables. However,
prior to the imposition of constraints, the states in ${\cal F}$ do not
necessarily describe physical states, and it does not have a good
meaning to ask what is the expectation value of an observable in such
a state.

     The rejection of the Hilbert space structure liberates us from a
straitjacket that often leads to inconsistencies \cite{K-ord}, but
it unfortunately leads to a loss of control over mathematical objects.
I shall later comment on both of these aspects.

\section{Space of solutions}

     The key idea of the Dirac constraint quantization is to turn the
constraints, which I shall now collectively call $\bf H$, into
operators (gate 2), and impose them as restrictions on the states:
                 \begin{equation}
{\bf H} \Psi = 0 \, .
                 \end{equation}
One surmises that only such states $\Psi$ which solve the constraints
can be physical. All physics is to be done on the space ${\cal F}_{0}$ of
states which solve Eq.(30).

     A number of remarks is appropriate. First of all, the quantum
constraints should not limit the quantum states more than the
classical constraints limit the classical states: they should not
beget other constraints by commutation.  This imposes stringent
requirements on the factor ordering of the constraint functions.  One
can see on simple models that these requirements virtually dictate the
factor ordering, and that to satisfy them the constraints cannot and
should not be represented by self-adjoint operators on ${\cal F}$
\cite{K-ord}. In geometrodynamics (and in connection dynamics), a
consistent factor ordering of constraints is a notorious unsolved
problem. The task is seriously hampered by the field-theoretical
aspects of canonical gravity, which call for regularization of the
constraint operators. \footnote{One should find a factor ordering of
the Hamiltonian and diffeomorphism constraints such that the
commutator of the Hamiltonian constraints yields an expression in which
the diffeomorphism constraint acts on the state function first,
followed by the structure functions of the `Dirac algebra'. It was
noticed by Anderson \cite{And} that this task cannot be accomplished
if one insists on representing the constraints by self-adjoint
operators on $\cal F$. A solution to the factor-ordering problem was
offered by Schwinger \cite{Schw} and criticized by Dirac \cite{Dir}.
The best, and certainly the shortest, exposition of Schwinger's
solution may be found in a footnote of the paper \cite{DW} by DeWitt;
this was later rediscovered by Komar \cite{Komar}. DeWitt himself made
a rather sweeping proposal on how to remove the problem by letting any
two field operators taken at the same point formally commute
\cite{DW}. Ashtekar \cite{Ash} proposed a simple factor ordering of
his constraints which (disregarding the regularization difficulties)
satisfies the consistency requirement. Unfortunately, all these results
are purely formal: Tsamis and Woodard \cite{T+W}, and Friedman and
Jack \cite{F+J} have persuasively argued that by formal manipulations
of the commutator one can obtain whatever result one wants.}

     The absence of a Hilbert space structure on ${\cal F}$ helps us
to make the constraints consistent. However, it also makes the
quantization badly dependent on the choice of representation. One can
see the problem already when solving the Schr\"{o}dinger equation of a
simple unconstrained system like an anharmonic oscillator
\cite{Ish-Sch},
                 \begin{equation}
\hat{h} = \case{1}{2} \hat{P} ^{2} + \case{1}{2} \hat{Q} ^{2} +
          \case{1}{4} \hat{Q} ^{4} .
                 \end{equation}
The solution of the Schr\"odinger equation calls for finding the
eigenfunctions of the energy operator (31). In the $Q$-representation,
the eigenfunction equation is a differential equation of the second
order. In the $P$-representation, it is a differential equation of the
fourth order. As a differential equation, the first equation has fewer
solutions than the second equation. The mismatch is removed by
requiring that the solutions we seek be square integrable (in $Q$ and
in $P$), i.e., belong to the Hilbert space based on the
Schr\"odinger norm.

     When, as in canonical gravity, we are unwilling to impose a
Hilbert-space structure on $\cal F$, the size of ${\cal F}$ depends on
the choice of representation. Thus, in principle, the solution space
$\Psi[g]$ in geometrodynamics is different from the solution space
$\Psi [p]$. Similarly, the connection representation $\Psi[A]$ is not
necessarily equivalent to the triad representation $\Psi [E]$.  In
other words, by not requiring that the representation space ${\cal F}
$ be a Hilbert space, one affirms a strong belief in the primacy of
those fundamental variables on which the representation is based.

     This is only a part of a larger problem. Unless one imposes some
boundary conditions on the solutions of the constraint equation (31),
the solution space ${\cal F}_{0}$ may be much too big. The quadratic
character of the Hamiltonian constraint in the momenta $p$ (or $E$)
evokes the analogy with the Klein-Gordon constraint for the
relativistic particle. There we know that the solution space of the
mass-shell constraint is also too big: the physical states of a
one-particle system correspond only to positive-energy solutions. In
geometrodynamics (and in connection dynamics), we do not have any
accepted method of cutting the basis of the solution space into half.
We are thus stuck with a solution space which may be physically too
big.

     If, on the other hand, we start imposing boundary conditions or
some other limitations on the states, we may inadvertently force the
solution space to be too small. This may (though it does not need
to) happen in connection dynamics when one requires that the states
$\Psi[A]$ be holomorphic functions of the complex connection $A$. One
imposes such a requirement in analogy with the Bargmann representation
for the states of a harmonic oscillator \cite{Barg}. The
solution of the eigenvalue equation for the oscillator Hamiltonian
$\hat{h}$ on the space of holomorphic functions of $Z = Q - i P$ gives
automatically a correct spectrum for $\hat{h}$. This is surprising,
because at this stage we do not yet have any Hilbert space. Only much
later is the space of holomorphic functions turned into a Hilbert
space which yields the same spectrum. This may not work so smoothly in
canonical gravity.  The complex connection is in some respects quite
different from the complex variable $Z$ for the harmonic oscillator.
(I shall return to this point three steps later.) There is a chance
that the `preestablished harmony' between holomorphic functions and
the subsequent construction of the Hilbert space no longer exists.

     To summarize, without the Hilbert space structure on ${\cal F}$
and without boundary conditions or some auxiliary conditions on the
states, we are bound to end up with a solution space ${\cal F}_{0}$
that contains many unphysical states. For this reason, I am reluctant
to call ${\cal F}_{0}$ `the physical space', and prefer to stick to a
more neutral name, the space of solutions.

\section{Observables}

     An outstanding question in the theory of constrained systems is
what dynamical variables can in principle be observed. An often made
proposal \cite{obs,Ash-book} is that
                 \begin{itemize}
                \item{Classical `observables' are those dynamical
variables $F$ whose Poisson brackets with the constraints weakly vanish:
                   \begin{equation}
{\bf H} = 0 \ \ \ \Longrightarrow \ \ \ \{ F,\, {\bf H} \} = 0 \, .
                   \end{equation} }
                 \end{itemize}
Its quantum mechanical counterpart is that
 \begin{itemize}
\item{Quantum `observables' are those operators $\hat{F}$ that commute
with the constraint operators $\hat{\bf H}$ on the space of solutions
{}${\cal F}_{0}\,$:

   \begin{equation}
\hat{\bf H} \Psi = 0 \ \ \ \Longrightarrow \ \ \ [\hat{F},\, \hat{\bf
H}]\Psi = 0\, .
   \end{equation}}
 \end{itemize}
     The second definition seems to be virtually forced on us if we
insist that the measurement of an observable does not throw the state
$\Psi$ out of the space of solutions ${\cal F}_{0}$.

     These two definitions are straightforward generalizations of the
concept of an observable in gauge theories. I am going to argue that
they are inappropriate for canonical gravity.

     To see why the definitions (32) and (33) are natural in ordinary
gauge theories, consider electrodynamics.  The vector potential $A$
describes the state of the electromagnetic field. The potentials
$A_{(1)}$ and $A_{(2)}$ which lie on the same orbit of the Gauss
constraint $G(x) = \nabla \cdot E(x)$ differ by a gauge transformation. They
are {\em physically indistinguishable}$\,$: they represent two equivalent
descriptions of the same physical state. One cannot observe the
individual $A$'s along the orbit, only the magnetic field $B$. The
magnetic field remains the same if we change the vector potential by a
gauge transformation:
 \begin{equation}
\{ B(x'),\, G(x) \} = 0 \, .
 \end{equation}
The magnetic field is an example of an observable.

     A quantum state of the electromagnetic field is described by the
state functional $\Psi [A]$. This functional is the probability
amplitude for finding the electromagnetic field in the state described
by the vector potential $A$. The probability should remain the same
when we change $A$ by a gauge transformation. This is ensured by the
Gauss constraint
 \begin{equation}
\hat{G}(x)\, \Psi [A] = 0 \, .
 \end{equation}
Equation (35) implies that $\Psi$ can depend on $A$ only via the
classical observable $B$: $\Psi = \Psi [B]$. On an ensemble of
systems described by the state functional $\Psi [B]$, we cannot
measure $\hat{A}$, but only $\hat{B}$. The magnetic field operator
{}$\hat{B}$ is a quantum observable. It satisfies the quantum
counterpart of Eq.(34):
 \begin{equation}
[\hat{B}(x'), \, \hat{G}(x) ] = 0 \, .
 \end{equation}

     The same case which I made for the Gauss constraint $G$ in
electrodynamics can be repeated for the rotation constraint $G_{i}$
and the diffeomorphism constraint $H_{a}$ in canonical gravity:

     Spacelike hypersurfaces in a Ricci-flat spacetime carry the
induced geometry, but do not come equipped with an orthonormal triad
$E$. The triad is a mere tool for calculating the metric (14). Two
triads, $E_{(1)}$ and $E_{(2)}\,$, on the same orbit of the constraint (16)
differ by a rotation. They both yield the same metric (14). Rotations
can be thought about as a gauge, and metric as an observable. In
general, the SO(3) observables are those dynamical variables which are
unaffected by rotations,
 \begin{equation}
G_{i}(x) = 0 \ \ \  \Longrightarrow \ \ \  \{ F, \, G_{i}(x) \} = 0 \, .
 \end{equation}

     In the triad representation, the quantum state of the
gravitational field is described by the state functional $\Psi [E]$.
The rotation constraint
 \begin{equation}
\hat{G}_{i}(x) \,  \Psi[E] = 0
 \end{equation}
implies that $\Psi$ can depend on $E$ only through the metric (14):
$\Psi = \Psi[g]$.

     However, the metric is not yet an observable with respect to
diffeomorphisms. Two metric fields, $g_{(1)}(x)$ and $g_{(2)}(x)$,
that differ only by the action of Diff$\Sigma$, i.e., which lie on the
same orbit of $H_{a}(x)$, are physically indistinguishable. This is
due to the fact that we have no direct way of observing the points $x
\in \Sigma$. A dynamical variable constructed from the metric field is
a true observable only if its value is unaffected by diffeomorphisms:
 \begin{equation}
H_{a}(x) = 0 \ \ \  \Longrightarrow \ \ \  \{ F,\, H_{a}(x) \} = 0 \, .
 \end{equation}
Thus, e.g., the volume of $\Sigma$ is an observable:
 \begin{equation}
V[g]  =  \int_{\Sigma} \d ^{3}x \, |{\rm det}(g(x))|^{\case{1}{2}} \, .
 \end{equation}

The momentum constraint
     \begin{equation}
\hat{H}_{a}(x) \,  \Psi[g] = 0
     \end{equation}
implies that the value of the state functional $\Psi[g]$ is the same
for all metrics connected by Diff$\Sigma$, i.e.\ , that $\Psi [g]$
does not depend on the individual metrics $g(x)$, but only on the
three-geometry ${}^{3}\cal G$.

     However, the definition (32) of an observable requires yet
something more. It claims that a dynamical variable $F$ cannot be
observed unless it has a vanishing Poisson bracket with the
Hamiltonian constraint $H$. I feel that this requirement is misguided.

     The action of $G_{i}$ on the dynamical variables generates their
change under rotations SO(3). The action of $H_{a}$ on the dynamical
variables generates their change under Diff$\Sigma$. Both of these
actions operate in the space of the instantaneous data on a fixed
hypersurface. The change of the data which they generate is
unobservable. The action of $H$ is different: it generates the
dynamical change of the data from one hypersurface to another. The
hypersurface itself is not directly observable, just as the points $x
\in \Sigma$ are not directly observable. However, the collection of
the canonical data $g_{(1)}$, $p_{(1)}$ on the first hypersurface is
clearly distinguishable from the collection $g_{(2)}$, $p_{(2)}$ of
the evolved data on the second hypersurface. If we could not
distinguish those two sets of the data, we would never be able to
observe dynamical evolution.

     The same reasoning applies to quantum theory. In the
Schr\"odinger picture, the evolution is carried by the state $\Psi$.
The Hamiltonian constraint
                            \begin{equation}
\hat{H}(x) \Psi = 0
                            \end{equation}
plays a different role from the diffeomorphism constraint or the
rotation constraint. It does not tell us that the evolved state is
indistinguishable from the initial state, but rather it tells us how
the state evolves. Thus, in geometrodynamics, the constraint (42) is a
second-order variational differential equation for the state $\Psi
[{}^{3}{\cal G}]$ of the three-geometry, called the Wheeler-DeWitt
equation \cite{Wheeler,DW}.
 This can be viewed as analogous to the
Klein-Gordon equation for the state $\psi (x^{\alpha})$ of a
relativistic particle. The three-geometry ${}^{3}{\cal G}$ is
considered as an internal configuration space{\/\em time} variable,
similar to the argument $x^{\alpha}$ of the Klein-Gordon state. The
Wheeler-DeWitt equation is supposed to describe the dynamical
evolution of the state in an internal configuration spacetime.

     It is this fundamental distinction between the states which are
and the states which are not distinguishable that leads me to reject
the definition (32) according to which `observables' should also have
a vanishing Poisson bracket with the Hamiltonian constraint. The
dynamical variable $F$ which satisfies this requirement,
                            \begin{equation}
H(x) = 0 \ \ \ \Longrightarrow \ \ \  \{ F, \, H(x) \} = 0 \,  ,
                            \end{equation}
must have the same value on all spacelike hypersurfaces. Therefore, it
is necessarily a constant of motion. This underscores the point which
I already made: If we could observe only constants of motion, we could
never observe any change.

     I hold that one can observe other dynamical variables,
like the volume variable (40), not only constants of motion.
Therefore, I shall call {\em observables} those dynamical variables
which are invariant under SO(3) and Diff$\Sigma$, but which do not
necessarily obey Eq.(43). Those observables which also satisfy Eq.(43) I
shall call {\em perennials\/}. I want to argue that
                            \begin{itemize}
   \item{One can observe dynamical variables which are not perennial,}
                            \end{itemize}
and that
                             \begin{itemize}
  \item{Perennials are often difficult to observe.}
                             \end{itemize}

     To make these two points, I do not need to deal with general
relativity. Any parametrized (or already parametrized) system
\cite{par} illustrates the same point. I shall try to clarify the
issues on the simplest of such systems, a parametrized free Newtonian
particle moving on a line. The phase space of the system is the
cotangent bundle $(T,Q; \,P_{T}, P)$ over the configuration spacetime
$(T,Q)$, and the Hamiltonian constraint amounts to the definition of
the energy $-P_{T}$ in terms of the momentum $P$:

                   \begin{equation}
H \, \mbox{:=} \, P_{T} + \case{1}{2} P^{2} = 0 \, .
                   \end{equation}

     Perform a canonical transformation \cite{Goa}
\begin{eqnarray}
Q' = Q - PT, &{}& P' = P, \\
T' = T,\ \ \ \ \ \ \ \ \       &{}& P_{T'} = P_{T} + \case{1}{2}P_{T}{}^{2}.
\end{eqnarray}
The primed canonical variables (45) are the initial data at $T=0$.
The primed time $T'$ is identical with the Newtonian time $T$. The
momentum $P_{T'}$ conjugate to $T'$ coincides with the Hamiltonian
constraint:
             \begin{equation}
H \, \mbox{:=} \, P_{T'} = 0 \, .
             \end{equation}

     Due to the constraint (47), any dynamical variable
$G(T',Q';\,P_{T'}, P')$ can be replaced by an equivalent variable
$F(T',Q';\,P') \,\mbox{:=} \, G(T',Q';\,0, P')$.  The variable $F$ is
a perennial if $\{F, \, H \} = 0\,$. Equation (47) enables us to
conclude that perennials are simply arbitrary functions of the initial
data. They cannot depend on $T'$:
             \begin{equation}
{\bf Perennials}\ :\ \ \ \  F = F( Q';\,P').
             \end{equation}
No perennial ever changes along a dynamical trajectory. To observe
change, we must observe at least one dynamical variable, like $T$ or
$Q$, which changes.

     An opposite view has been expressed by Rovelli \cite{Rovelli}. I
interpret his paper as saying that to observe a changing dynamical
variable, like $Q$, amounts to observing a one-parameter family
             \begin{equation}
Q'(\tau ) \, \mbox{:=} \, Q' + P' \, \tau = Q - P \,(T-\tau ),
\ \ \tau \in {\sf R}
             \end{equation}
of perennials. The perennials (49) are the values of $Q$ at $T=\tau$.
By observing the perennials $Q'(\tau _{1})$ and $Q'(\tau _{2})$ one
can infer the change of $Q$ from $T= \tau _{1}$ to $T = \tau _{2}\,$.

     The problem with such a view is that one is not told how to
observe $\tau$. One way of observing $\tau$ is to watch the dynamical
variable $T$ (the hand of an ideal Newtonian clock). The value of $T$
is $\tau$. However, this amounts to observing a dynamical variable $T$
which is not a perennial. An alternative is to say that one can
observe $\tau$ directly. Again, one is forced to admit that one can
observe an entity which is not a perennial. The third alternative is
to say that because perennials are constants of motion, it does not
matter when they are observed. One can observe all the perennials
$Q'(\tau)\,, \ \tau \in {\sf R}$ at once, and infer `the change of $Q$
with $T$' from that instantaneous observation.  Any instant is like
any other, and each contains the same set $\tau \in {\sf R}$ of
perennials from which the change is inferred.  This does not make me
too happy either. If all time $\tau$ is eternally present, all time is
irredeemable.

     My discussion was so far concerned with the epistemological
status of observables. I tried to argue that the identification of
observables with perennials drives one to a Parmenidean view of the
world. Physicists are soundly sceptical of epistemological arguments,
and I am not deluding myself that my argument is an exception.
``Refutations are seldom final; in most cases, they are only a prelude
to further refinements.'' Significantly, Bertrand Russell made this
remark when closing his discussion of Parmenides \cite{Russell}.

     So far I argued that some observables are not perennial. I
must now defend my other point, namely, that perennials are often
difficult to observe. In this part of the discussion, I take the
attitude of physical common sense, that at any instant one can
directly observe the position $Q$ of the particle, its momentum $P$,
and the time $T$ on an ideal Newtonian clock, but not the position
$Q'$ which the particle had at time $T = 0$. The initial position
$Q'$, which does not change with T and is a perennial, is {\em
inferred} from the observed data $Q$, $P$, and $T$ by using Eq.(45).
For a free particle, such an inference is easy because we know how to
integrate equations of motion. However, even for such a simple system
as a free particle, the inference may be hampered by experimental
errors. If one determines $P$ with an error $\Delta P$, the error in
the inferred value of $Q'$ scales with $T$. If the particle moves on a
circle and T is large, it is practically impossible to infer from the
observations at T where on the circle the particle was at $T=0$. For
more complicated Hamiltonians, like those governing dynamics of many
interacting particles, the task of inferring perennials becomes pretty
hopeless. Take, e.g., a globular cluster, observe the current
positions and momenta of the stars, and then try to infer what were
their positions and momenta when the cluster was formed some 15
billion years ago.

     In quantum theory, there is yet another reason why perennials are
difficult to observe. To measure a quantum variable, one needs to
design an apparatus with appropriate coupling. \footnote{My discussion
takes place within the framework of von Neumann's theory of
measurement. It should be rephrased in a scheme like that advocated by
Hartle in the present volume.} Theoretically, it is possible to find
an apparatus which measures an arbitrary quantum variable $\hat{F} =
F(\hat{Q} , \hat{P})\,$. Experimentally, this can be done only for a
small number of especially simple variables, like $\hat{F}=\hat{Q}$ or
$\hat{F}=\hat{P}$. For elementary systems, like a free particle or a
harmonic oscillator, the initial-data perennials (45) are linear
functions of the current data $\hat{Q}$ and $\hat{P}$.
Experimentalists know how to build the apparatuses for measuring such
perennials. An example is the discussion of non-demolition experiments
for detecting gravitational waves \cite{non-demol}. To circumvent the
limits imposed by the uncertainty principle, one constructs an
apparatus for monitoring the initial length of an oscillating bar,
i.e., the perennial like $Q'$ of Eq.(45) for a linear harmonic
oscillator.  However, even for such a simple system as the hydrogen
atom, the initial-data perennials are complicated functions of the
current data $\hat{Q}$ and $\hat{P}$. It is difficult to conceive an
apparatus which would monitor such perennials at all times.

     If the dynamical system is not Newtonian, i.e., if the
Hamiltonian constraint is not linear in the momentum $P_{T}$ conjugate
to a time variable $T$, the practical difficulty of determining
classical perennials from the current data turns into something much
more serious: into an argument questioning their very existence. A
classical example is an asymmetric top spinning around a fixed point
in a homogeneous gravitational field. Describe the configuration of
the top by the Euler angles $Q^{a}=(\phi, \psi, \theta)$, where
$\theta$ is measured from the direction of the field. The
Hamiltonian $h$ of the top is a quadratic function of the momentum
$P_{a}$. Constrain the motion of the top to be taking place with a
definite energy $E$:
             \begin{equation}
H \, \mbox{:=} \, h-E=0 \,,\ \ h=\case{1}{2}G^{ab}(Q)P_{a}P_{b} + V(Q) \, .
             \end{equation}
The trajectory of the top in the phase space $(Q^{a},P_{a})$ is
generated by the Hamiltonian constraint (50). Notice that we do not
ask how the top {\em moves} in the Newtonian time $T$, we are merely
asking about its {\em trajectory}. The momentum $P_{T}$ does not enter
into the constraint (50); it was replaced by a constant $E$.

     Perennials are defined as those dynamical variables
$F(Q^{a},P_{a})$ that have a (weakly) vanishing Poisson bracket with
$H$. Notice that $T$ cannot be used in the construction of perennials
because it no longer is a canonical variable.

      One perennial is the angular momentum $M_{\theta}$ about the
direction of the gravitational field. This perennial is linear in the
momentum $P_{a}$:
             \begin{equation}
M_{\theta} = M^{a}(Q)P_{a}\, .
             \end{equation}
A century ago, Poincar\'e asked the question \cite{Poincare}: Does the
top have any other integrals of motion than those of vis viva and the
area? In the way I formulated the problem, this translates into the
question: Is there any perennial besides $M_{\theta}\,$? The answer
is {\em no\/} \cite{Arnold}.

     The configuration space $(T,Q)$ of a parametrized free Newtonian
particle is two-dimensional, and there is one Hamiltonian constraint.
There are $2 \times (2-1) = 2$ independent perennials (45); any other
perennial is their function. An $n$-dimensional parametrized Newtonian
system should have $2(n-1)$ independent perennials. The top is a
three-dimensional system, and one would expect to find four
independent perennials. However, the constraint (50) does not have the
Newtonian form, and there is only one perennial, (51).

     Let me briefly return from models to canonical gravity. General
relativity is not a parametrized field theory whose constraints have a
`Newtonian' form (44). In particular, both in geometrodynamics and in
connection dynamics, the Hamiltonian constraint is quadratic in the
momenta. The supermetric has some non-trivial dependence on the
canonical coordinates. In these respects, the Hamiltonian constraint
resembles the constraint (50) for the top. This prompts the following
remarks:
             \begin{itemize}
               \item {We do not know how to construct perennials for
                     canonical gravity.}
               \item {We do not know how to select families of
                     perennials (similar to the family (49)) labeled
by a functional time parameter (similar to $\tau$) which would
correspond to `simple' dynamical variables as the volume observable
(40) (similar to $Q$).}

               \item {So far, we did not find a single gravitational
                     perennial. \footnote{It is not clear whether the
interesting result reported at this meeting by Goldberg {\em et al.\/}
can be recast into a construction of a perennial.} The existence of a
complete set of perennials would imply that gravity is a completely
integrable theory.  They are indications that it is not
\cite{K-nosym,And+Torre}. It is likely that the gravitational
perennials are rare, and it is quite possible that there are none.}
             \end{itemize}

     Perennials in canonical gravity may have the same ontological
status as unicorns ---{\em a priori\/}, these are possible animals,
but {\em a posteriori\/}, they are not roaming on the Earth. According
to bestiaries, the unicorn is a beast of fabulous swiftness, strength,
and beauty, but, alas, it can be captured only by a virgin
\cite{Warner}.  Corrupt as we are, we better stop hunting mythical
beasts.

\section{Hilbert space}

     Once we have decided what dynamical variables can be observed, we
need to know what is the statistical distribution of their observed
values. In quantum mechanics, probabilities are determined by the
inner product in a Hilbert space. Therefore, we need to endow the
space of physical states with a Hilbert space structure.

     The proposals on how to find the inner product depend on what
position one takes on observables. Let me first discuss the proposal
\cite{Ash-qp}, which relies on identifying observables with
perennials:
\begin{itemize}
\item{Choose an inner product $ \langle \Psi _{1}| \Psi _{2} \rangle$
on the solution space ${\cal F}_{0}$ such that all {\em real} quantum
perennials are self-adjoint under it.}
\end{itemize}
     In geometrodynamics, the phase space is real and it is easy to
say when a dynamical variable is real. I return to the reality problem
in connection dynamics in the next section.

     There are several problems with the above proposal.  First of
all, we have seen that there may not be any perennials in canonical
gravity, or that at least there may not be a sufficient number (a
complete set) of them. If so, the proposal on how to determine the
inner product either loses its content, or becomes too weak.
Secondly, even when one disregards this difficulty, one should notice
that the proposal as it stands is self-contradictory. If $\hat{F}$ and
$\hat{G}$ are quantum perennials, so is $\hat{F} \hat{G}$. If
$\hat{F}$ and $\hat{G}$ are self-adjoint under the inner product
$\langle \Psi _{1} | \Psi_{2} \rangle$, $ \hat{F} \hat{G}$ is not. To
remove the contradiction, one needs to find `fundamental perennials',
and approximate all other perennials by polynomials of the fundamental
perennials. One can then require that only the fundamental perennials
be self-adjoint, and symmetrically factor order the polynomials which
define the remaining perennials. Unfortunately, the original
fundamental variables $g$ and $p$ (or $A$ and $E$) are not perennials,
and we lack a guiding principle on what the fundamental perennials may
be.

     The third problem with the proposal is that the solution space
${\cal F}_{0}$ is probably larger than the space of physical states.
We have seen that it may contain `improper elements', `unbounded
states', and `states with negative norms'. The definition of a
perennial $\hat{ F}$ requires that $\hat{ F}$ commutes with the
constraints on the solution space ${\cal F}_{0}$. If ${\cal F}_{0}$ is
too large, the set of perennials may be too small: Some physically
significant perennials may have been excluded by the requirement that
they commute with the constraints on a larger-than-physical space of
solutions. Further, it may happen that those perennials which remain
cannot be made self-adjoint under an inner product on the whole
solution space, but only on a drastically reduced space from which the
`unphysical' states have been excluded. In brief, it seems impossible
to follow step by step the `quantization program': firstly, to find
the space of solutions without having the inner product to determine
which states are physical, secondly, on that space of solutions to
define the perennials, and thirdly, to find the inner product on
${\cal F}_{0}$ which makes all such perennials self-adjoint. Rather,
the steps should be replaced by a single jump. As I am growing older,
the difficulty of replacing three steps by a single jump is becoming
more and more obvious.

     The second standpoint is that observables do not need to commute
with the Hamiltonian constraint, but only with the gauge constraints.
If so, they do not act in the space of solutions: if $\Psi \in {\cal
F}_{0}$ and $\hat{F}$ is an observable, $\hat{F} \Psi \notin {\cal
F}_{0}$. To proceed, one should
\begin{itemize}
\item{abandon the space of solutions and work
instead in the space of instantaneous states.}
\end{itemize}

     To talk about instantaneous states requires a decision about
what is an instant. An instant in a relativistic spacetime is a
spacelike hypersurface. However, spacelike hypersurfaces are not
elements of the gravitational phase space. The task is to find an
observable $T$ (or, rather, a set of $\infty ^{3}$ commuting
observables, to account for $\infty ^{3}$ hypersurfaces) whose value
uniquely fixes a hypersurface in a Ricci-flat spacetime generated by
the evolution of the classical canonical data. Such an observable is
called an {\em internal time}. (The adjective `internal' means
`constructed solely from the phase-space variables'.)

     The Hamiltonian constraint is interpreted as an evolution
equation for $\Psi$ in $T$. One tries to cut down ${\cal F}_{0}$ to a
linear subspace $ {\cal F}_{0}' \subset {\cal F}_{0}$ whose elements
are in a one-to-one correspondence with the instantaneous values of
$\Psi$: the restrictions $\Psi _{T}$ of $\Psi$ to a fixed hypersurface
$T$. These restrictions are the instantaneous states $\Psi _{T} \in
{\cal F}_{T}$. The program is to find an inner product in ${\cal
F}_{T}$ which is independent of $T$, i.e., which is conserved in
internal time. The discussion centers on how different forms of the
Hamiltonian constraint (the Wheeler-DeWitt form, and others) suggest
what such an inner product may be. A $T$-independent inner product can
be interpreted as an inner product in ${\cal F}_{0}' \,$. In general,
the observables $\hat{F}$ depend on $T$. One requires that they be
factor ordered so that, at each $T$, they are self-adjoint under the
inner product in ${\cal F}_{T} \,$. The expression $ \langle \Psi _{T}
| \hat{F} | \Psi _{T} \rangle$ is interpreted as the mean value of
$\hat{F}$ in the state $\Psi _{T}$ at the internal time $T$.

     These things are more easily said than done. The internal time
proposal meets as many difficulties as the approach based on the
concept of perennials.  I discussed the problems of time in a recent
review \cite{K on t} which complements my present treatment
of observables.

     It is sometimes maintained that the approach based on perennials
somehow avoids the problems of time. It would be great if it did, but
I fear it does not. A closer look reveals that the problems of time
and the problem of perennials are rather closely related. A Czech
saying has it that the devil thrown out of the door returns through a
window.

\section{Reality conditions}

     The connection dynamics looks in many respects simpler than
geometrodynamics, but its simplicity has been bought at a price: the
SO(3) connection $A$ is necessarily complex. One needs to ensure that
the quantum theory based on such a connection describes a real
gravitational field.

     One can attempt to accomodate complex objects in canonical
gravity in two different ways:

     {\em Complexify the Einstein theory\/}, i.e., work with
complex metrics $\gamma$ on a real spacetime manifold $\cal M$. The
statement that $({\cal M}, \gamma)$ is Ricci-flat amounts to a system
of coupled equations for the real and imaginary parts of the complex
metric $\gamma\,$. These equations can be derived from a real action
whose Lagrangian is the real part of the complex curvature scalar.
Introduce the Ashtekar variables $A^{i}_{a}\,$, $E^{a}_{i}$ for the
complexified spacetime. Both $A$ and $E$ are now complex. The
canonical form of the action leads to the Poisson brackets among these
variables and their complex conjugates.

     To restrict the spacetime metric to be real, one imposes the
condition that its imaginary part vanishes. In the canonical version
of the theory, this imposes the reality conditions (27) on $A$ and
$E$. The reality conditions are preserved by the constraints: when the
evolution starts from real canonical data, it continues building a
real spacetime. However, the Poisson brackets among the reality
conditions do not vanish: to put the imaginary part of the metric and
its rate of change equal to zero amounts to requiring both a canonical
coordinate and its conjugate momentum to vanish. It means that the
reality conditions are, in Dirac's terminology, second-class
constraints \cite{Dirac}.  Such constraints must be eliminated before
quantization.  Unfortunately, their elimination destroys the new
variables.

     An alternative is to derive the complexified equations from a
holomorphic Lagrangian \cite{Jacobson}. The corresponding canonical
theory knows how to form the Poisson brackets among $A$ and $E$, but
the Poisson brackets involving the complex conjugates $\bar{A}$ and
$\bar{E}$ are undefined. The status of the reality conditions thus
remains unclear and one does not know what to do with them on
quantization.

     {\em Use complex chart on a real phase space.} The second option
is to consider $A$ and $E$ as a complex chart on a real phase space
($E,\,-K$). This is similar to introducing a complex chart $Q$ and
$Z=Q-iP$ on the real phase space ($Q,\,P$) of a harmonic oscillator.
The proposal \cite{Ash-book,A+T} is to ignore the reality conditions
in the first five steps of the quantization program. In particular,
the vector space $\cal V$ spanned by the fundamental variables $A$ and
$E$ is allowed to be complex, and so are the dynamical variables
$F[A,E] \in \cal A$ and the perennials $F[A,E] \in {\cal A}_{0}\,$.

     One knows how to complex conjugate, $\bar{F}\,$, the
elements $F$ of the classical spaces $\cal V$ and $\cal A$. The task
is to define the corresponding operation, ${\star}\,$, on the
elements $\hat{F}$ in $\hat{\cal V}$ and $\hat{\cal A}$. Ashtekar's
proposal is first to define the $\star$ operation in $\hat{\cal V}$ by
requiring that complex conjugate elements of $\cal V$ are carried into
the $\star \,$- related elements of $\hat{\cal V}\,$:
\begin{equation}
F,\bar{F} \in {\cal V} \ \ \Longrightarrow \ \
\widehat{F}=\widehat{\bar{F}}{}^{\star}\, .
\end{equation}
The $\star$ operation is then extended from $\cal V$ to $\cal A$ by
using the axioms of the involution operation:
\begin{eqnarray}
(a\hat{F} + b\hat{G})^{\star} &=& \bar{a}\hat{F}^{\star}
                                  + \bar{b}\hat{G}^{\star}\, ,
                                  \nonumber \\ ~
(\hat{F}\hat{G})^{\star}      &=& \hat{G}^{\star}\hat{F}^{\star}\,, \\
(\hat{F}^{\star})^{\star}     &=& \hat{F}\, , \nonumber \\
\forall \hat{F},\hat{G} \in {\cal A} &{\rm and}& \forall a,b \in {\sf
C}\, .                                   \nonumber
\end{eqnarray}

     If $\hat{F}^{\star} = \hat{F}\,$, the operator $\hat{F}$
represents a real dynamical variable. If there are no constraints,
this dynamical variable is an observable. The expectation value of
$\hat{F}$ should be real. This objective can be achieved by requiring
that the inner product $\langle \Psi _{1} | \Psi _{2} \rangle$ in
$\cal F$ be such that it makes all $\star \,$-related operators
Hermitian adjoints,
\begin{equation}
\hat{F}=\hat{G}^{\star} \ \ \Longrightarrow \ \ \langle \Psi _{1}
       | \hat{F} \Psi _{2} \rangle = \langle \hat{G}
       \Psi_{1}|\Psi_{2}\rangle \, ,
\end{equation}
and hence all operators representing real variables self-adjoint. If
the $\star$ operation in $\hat{\cal A}$ is determined by the $\star$
operation in $\hat{\cal V}$ as in Eqs.(52) and (53), it is sufficient
to require that the condition (54) holds for all fundamental variables
$\hat{F} ,\hat{G} \in \hat{\cal V}\,$.

     Canonical gravity, however, is a constrained system. Ashtekar's
program assumes that only perennials can be observed, and that their
expectation values are obtained from an inner product on the space of
solutions. To impose the reality conditions, one needs to define the
$\star$ operation for perennials. This would be straightforward if the
$\star$ operation from $\hat{\cal A}$ could be restricted to
perennials. Unfortunately, this does not need to be the case: if
$\hat{F}$ is a perennial, $\hat{F}^{\star}$ does not need to be a
perennial (though it may be a perennial under special circumstances).
The hope is that there is a `sufficient' number of perennials
$\hat{F}$ whose $\star\,$-adjoints $\hat{F}^{\star}$ are also
perennials. By `sufficient' one means that the condition (54), when
imposed on these perennials, uniquely determines the inner product in
${\cal F}_{0}\,$.

     To summarize, Ashtekar's program calls for implementing the
reality conditions as requirements on the inner product in the space
of solutions ${\cal F}_{0}\,$. Firstly, one must find a sufficient
number of $\star \,$-adjoint perennials, and then require that these
be Hermitian adjoints under the inner product.

     One can ask two questions about this proposal. The first is
whether it works for simple model systems. The second is whether it
can reasonably be expected to work in canonical gravity.

     The answer to the first question is yes. Ashtekar's proposal
determines the inner product for a number of simple systems (a
harmonic oscillator with complex chart, a parametrized Newtonian
particle, a free relativistic particle on a flat background). It also
works for $2+1$ gravity and linear field theories on a
$(3+1)$-dimensional flat Lorentzian background, including Maxwell's
electrodynamics and linearized gravity. With the exception of $2+1$
gravity (which does not have any field degrees of freedom) these
examples are reducible to collections of harmonic oscillators.

     To approach the second question, one should ask whether there are
any relevant differences between the prototype of a linear harmonic
oscillator and full canonical gravity. (By `relevant' I mean relevant
to the proposal on handling the reality conditions.) I feel there are
two such differences:

     In the harmonic oscillator problem, one works with the
fundamental variables $Q$ and $Z = Q - iP\,$, which are analogous to
$E$ and $A$ in connection dynamics. The vector space $\cal V$ is
spanned on $Q$, $Z$, and 1. The reality conditions are the conditions
\begin{equation}
\bar{Q} = Q \ \ {\rm and}\ \ \bar{Z} = -Z + 2Q
\end{equation}
on the dynamical variables $F=Q$, $G=Z$, and their complex conjugates
$\bar{F}$ and $\bar{G}\,$. Both $F$ and $G$, and $\bar{F}$ and
$\bar{G}$ lie in $\cal V \,$. It is thus possible to define the
$\star$ on $\cal A$ by Eqs.(52) and (53), and to impose the reality
condition (54).

     In connection dynamics, $\cal V$ is spanned by $A$, $E$ and 1.
The second reality condition (27), however, is not a condition on the
elements of $\cal V$, because $\Gamma [E]$ is a non-linear functional
of $E$. (The same remark applies to polynomial forms of reality
conditions.) This prevents one from defining the $\star$ operation on
$\cal V$, as in Eq.(52), and from extending it to $\cal A$, as in
Eq.(53).

     This difficulty can be clarified on simple models. Take a
one-dimensional system with the Hamiltonian \footnote{In the sector
$Q>0$, the Hamiltonian (56) can be brought into the form $h = \case
{1}{2} p^{2} + 4 q^{-4}$ by the canonical transformation $q = \sqrt{2}
\, Q^{\case{1}{2}}\,$, $p = \sqrt{2}\, Q^{\case{1}{2}}P\,$.}
\begin{equation}
h\,\mbox{:=}\, QP^{2} + Q^{-2}
\end{equation}
and introduce the complex chart $(Q,Z)$, with
\begin{equation}
Z = Q^{-1} - iP,
\end{equation}
on the real phase space $(Q,P)$. The Hamiltonian (56) becomes
polynomial in $Q$ and $Z$,
\begin{equation}
h = -QZ^{2} + 2Z.
\end{equation}
The reality condition on Z can be written either in a non-polynomial
form linear in $Z$, or in a polynomial form:
\begin{equation}
\case{1}{2}
(Z + \bar{Z}) = Q^{-1}\,,\ \ \ {\rm or}\ \ \  Q(Z+\hat{Z})=2 \, .
\end{equation}
Whichever form we use, it is not a condition in the complex vector
space $\cal V$ spanned by the fundamental variables $Q$ and $Z$.

     This is my first reason for believing that the harmonic
oscillator is not quite representative of canonical gravity. The
algorithm for handling reality conditions needs to be checked on more
general models than those which have been investigated so far, like
the model I have just described.

     The second difference between the oscillator and canonical
gravity is that the later is a parametrized theory. Ashtekar's
proposal on how to handle the reality conditions depends on the
existence of a sufficient number of perennials, and on the possibility
to define a $\star$ operation on their algebra. I expressed my doubts
that there exists a sufficient number of perennials in canonical
gravity. Even if there is a sufficient number of perennials, it
remains unclear whether it is possible to extend the $\star$ operation
{}from $\hat{\cal V}$ to $\hat{\cal A}\,$, and then to restrict it to a
suitable subset of perennials.

 I do not claim that these problems are insurmountable, but I feel
that they represent a major unsolved problem of connection dynamics.

\section{Conclusions}

     Where do we stand? We certainly gained in the years a good
geometric understanding of classical general relativity as a canonical
dynamical system. In quantum theory, we inherited a set of rules of
thumb called Dirac constraint quantization. They were never precise,
and Dirac himself never claimed they were much more than rules of
thumb. People tried to make them more precise and they ended with
something resembling the seven gates I described.

     Let me revisit those gates and ask what steps in the quantization
program have actually been accomplished. And, even more importantly,
let me summarize what are the main unsolved problems.
\begin{enumerate}
  \item{Different sets of fundamental variables (not only those which
        I mentioned in this report) have been explored and understood.
We also know how to take care of the positivity restrictions on the
metric variables \cite{Ish+}}.
  \item{Most of the work on turning constraints into operators is
        formal. Both the regularization problem and consistency
problem remain open. Very little is known about how to handle other
dynamical variables, especially the future candidates for observables
or perennials.}
  \item{Important work has been done on clarifying the mathematical
        status of the states $\Psi[g]$ and $\Psi[A]$ and of the
fundamental operators \cite{Ish+}. The connection representation
has been linked to the loop representation \cite{A+I}. One should note
that the latter investigation has been successful only for {\em
real\/} connections.}
  \item{Because the regularization and consistency problems for the
        constraints have not been satisfactorily resolved, all
attempts to find the states which solve the quantum constraints (30)
are to a large extent formal. It is notable that connection dynamics
actually exhibited a large number (indeed, infinitely many) such
solutions. Most of these were obtained in the loop representation and
lie outside the scope of this report \cite{Rov-loops}. When comparing
this success with the lack of solutions in geometrodynamics, one
should keep in mind that these solutions correspond to degenerate
metrics which geometrodynamics excludes. One solution that can be
written directly in the connection representation is the exponential
of the Chern-Simons form \cite{Kodama}. Passing from particular
solutions to general considerations, it is not clear what boundary or
other conditions should be imposed on the solutions $\Psi \in {\cal
F}_{0}$ to select the true physical states.}
\item{The problem of what quantities can be observed (and how they can
      be observed) is one of the most intriguing and important
questions in quantum gravity. A widely held view (which I dispute) is
that one can observe only perennials. No true perennials, classical or
quantum, have so far been found, and even if they exist, finding them
is difficult. I feel we should instead concentrate on formulating and
proving (non?)existence theorems about perennials. \\
     Unlike perennials, there are many concrete examples of classical
observables. It is, however, obscure what classical observables are to
be represented by operators, and on what space these operators act.
This is connected with the problem of time: one does not expect the
time observable to be represented in quantum mechanics by an operator.}
\item{Another outstanding problem of canonical quantum gravity is the
      construction of the inner product. Quantum geometrodynamics has
been unsuccessful in this task \cite{DW,K-nosym}, and connection
dynamics has hardly done more than formulate broad guidelines on how
one might try to proceed. These guidelines crucially depend on the
existence of perennials. \\ In contrast, one knows how to construct
(at the formal level) the inner product for parametrized field
theories \cite{K-Ox}. Each choice of an internal time casts
canonical gravity into the mold of a parametrized field theory and
leads to an inner product. The procedure, however, is not without
problems \cite{K on t}. One which is closely related to the
problem of perennials is that internal time may not exist globally
 \cite{Torre-QGnotP}}.
\item{Connection dynamics, unlike geometrodynamics, needs to take care
      of reality conditions. Ashtekar's proposal is to impose them
as requirements which determine the inner product.
Two problems arise: firstly, the necessity of finding a complete set
of perennials and defining on them the $\star$ operation and,
secondly, the high polynomiality of the reality conditions, which
takes them out of the realm of the fundamental vector space $\cal
V$.\\ The reality conditions are the only major problem which does not
exist in geometrodynamics. The ability of connection dynamics to
handle this problem will be
 crucial for judging its success in the
Galilean contest between the two chief systems of canonical gravity.}
\end{enumerate}

     The problem of reality conditions exemplifies the general pitfall
of any quantization program. As I described it, the program resembles
seven doors to the law, each of them guarded by a doorkeeper. We
certainly did not sit on a stool at the side of the first door for
days and years: we tried to enter the law. However, on our way through
the doors we learned that their orderly sequence is deceptive. One can
never be sure of passing a door before all have been passed. The
entries are so interconnected that they cannot be made separately:
What is a solution of the quantum constraints depends on the choice of
fundamental variables and the form of the constraints. What solutions
are physical depends on the inner product. What is an inner product
depends on what quantities are observable. What quantities are
observable may depend on what solutions are physical. More often than
not we are caught in a vicious circle which calls for entering all the
doors at once.

     This may be frustrating, but it should have been expected.
Indeed, it would be rather disappointing if one could reach a truly
fundamental theory like quantum gravity by following step by step a
travel guide, or its medieval predecessor, a pilgrim's itinerary to a
wholy shrine. In this spirit, let me end my account of canonical
quantum gravity in the dark ais
les of St. V\'{\i}t's cathedral of my native
city of Prague, talking to a priest \cite{Kafka}:
\begin{quotation}
``You have studied the story more exactly and for a longer time than I
have,'' said K. They were both silent for a while. Then K. said: ``So
you think that the man was not deceived?'' ``Don't misunderstand me,''
said the priest, ``I am only showing you the various opinions
concerning that point. You must not pay too much attention to them.
The scriptures are unalterable and the comments often enough merely
express the commentators' despair.''
\end{quotation}

\section*{Acknowledgments}
The work on this report has been partially supported by the NSF grants
PHY-9207225 and INT-8901512 to the University of Utah. I want to thank
Julian Barbour for his careful reading of the final draft of the paper.

\end{document}